\documentclass[AMA,STIX2COL,Linenumbersoff]{MRM}
\articletype{Note}%

\usepackage{mathtools}
\usepackage{amsmath}

\usepackage{optidef}

\received{TBD}
\revised{N/A}
\accepted{N/A}
\begin{document}

\title{Variable Resolution Sampling and Deep Learning-Based Image Recovery for Faster Multi-Spectral Imaging Near Metal Implants}

\author[1]{Nikolai J. Mickevicius}{}
\author[1]{Azadeh Sharafi}{}
\author[1]{Andrew S. Nencka}{}
\author[1]{Kevin M. Koch}{}

\authormark{Mickevicius \textsc{et al}}

\address[1]{\orgdiv{Department of Radiology}, \orgname{Medical College of Wisconsin}, \orgaddress{\city{Milwaukee}, \state{Wisconsin}, \country{USA}}}

\corres{Nikolai Mickevicius, Department of Radiology, Medical College of Wisconsin, 8701 Watertown Plank Road, Milwaukee, WI, 53226. \email{nmickevicius@mcw.edu}}



\abstract{\section{Purpose} In multi-spectral imaging (MSI), several fast spin echo volumes with discrete Larmor frequency offsets are acquired in an interleaved fashion with multiple concatenations. Here, a variable resolution (VR) method to nearly halve scan time is proposed by only acquiring low resolution autocalibrating signal in half of the concatenations.  

\section{Methods} Knee MSI datasets were retrospectively undersampled with the proposed variable resolution sampling scheme. A U-Net model was trained to predict the full-resolution images from the VR input. Image quality was assessed in 10 test subjects. 

\section{Results} Spectral bin-combined images produced with the proposed variable resolution sampling with deep learning reconstruction appear to be of high quality and exhibited a median structural image similarity of 0.984 across test subjects and slices. 

\section{Conclusion} The proposed variable resolution sampling method shows promise for drastically reducing the time it takes to collect multi-spectral imaging data near metallic implants. Further studies will rigorously examine its clinical utility across multiple implant scenarios. 

} 



\maketitle

\footnotetext{\textbf{Abbreviations:} MSI, multi-spectral imaging;}

\clearpage
\section{Introduction}\label{sec:intro}

High-susceptibility metallic implants induce large perturbations in the $B_0$ field\cite{koch2010magnetic,hargreaves2011metal}. While the use of fast spin echo (FSE) pulse sequences mitigates signal dropout due to intravoxel dephasing, the resulting images remain corrupted by severe distortion and missing signal from spins resonating outside of the radiofrequency bandwidth. Multi-spectral imaging (MSI)\cite{koch2009multispectral,koch2011imaging} was developed to overcome these obstacles by acquiring several (e.g, up to 24) band-limited FSE volumes at discrete Larmor frequency offsets. These so-called spectral "bin" images are then combined after reconstruction to yield an artifact-reduced volume. 

While MSI produces high quality images in close proximity to metal implants, it comes as the expense of long scan times owing to the need to acquire many 3D FSE volumes. Acquiring only a subset of bin volumes within each TR period (i.e., using concatenations) is performed to minimize partial saturation artifacts between overlapping spectral profiles\cite{koch2015flexible}, thereby demanding scan times of TR $\times N_s \times N_c$. Here, $N_s$ is the number of shots, or echo trains, needed to fill k-space, and $N_c$ is the number of concatenations. 

For typical MSI-based proton density weighted imaging of the knee, $N_c=2$, TR $\approx 4$ s, and $N_s \approx 40$, leading to scan times of approximately 5.5 minutes. The approximated number of shots (40) needed to fill k-space is typical for phase encoded matrix sizes of 256x32 with an echo train length of 32, 2x2 acceleration and partial Fourier sampling with 8 oversampling lines.   Higher numbers of phase encodes with modest increases in acquisition time, enabling isotropic MSI-acquisitions~\cite{zochowski2019mri}, can be attained using longer echo trains with refocusing flip-angle modulation and by reducing the number of acquired spectral bins through prospective calibration procedures~\cite{kaushik2016external}.  These isotropic applications remain limited in scope, however, due to insufficient isotropic resolution and/or unacceptably long acquisition times.  

With acceleration already being pushed to the limit of conventional parallel imaging methods, further accelerating MSI to improve its spatial resolution capabilities while maintaining patient comfort and compliance is non-trivial.  

Work in the area of non-linear model-based image reconstruction has been proposed to dramatically decrease scan times while vastly improving image quality over parallel imaging+compressed sensing algorithms\cite{shi2019accelerated}. This approach, however, requires extensive reconstruction times, even with the use of a high-performance reconstruction server.   In the present work, a variable resolution (VR) sampling scheme and deep learning reconstruction framework is proposed to shift the computational burden to offline training of a model that can be rapidly applied in a clinical environment. The VR sampling scheme acquires routinely utilized 2x2 and partial Fourier accelerated k-space data in half of the concatenations.  Only the parallel imaging autocalibrating signal (ACS) k-space lines\cite{griswold2002generalized} are acquired in the other half of the concatenations. Thus, the goal of this work is to reduce the scan time to TR $\times \frac{N_c}{2} \times (N_s + N_{s,acs})$, where $N_{s,acs}$ is the number of shots needed to acquire the ACS data only. This approaches a factor of two times faster than conventional MSI when $N_{s,acs} << N_s$. 

From a mathematical perspective, this is a feasible approach since the smoothly varying spectrally selective profiles in each bin overlap significantly between adjacent bins. Using this relationship, a deep learning model is proposed to take take the VR bin images as input to predict full-resolution images from the bins reconstructed from ACS data only. 

\begin{figure}[ht]
\centerline{\includegraphics[width=21pc]{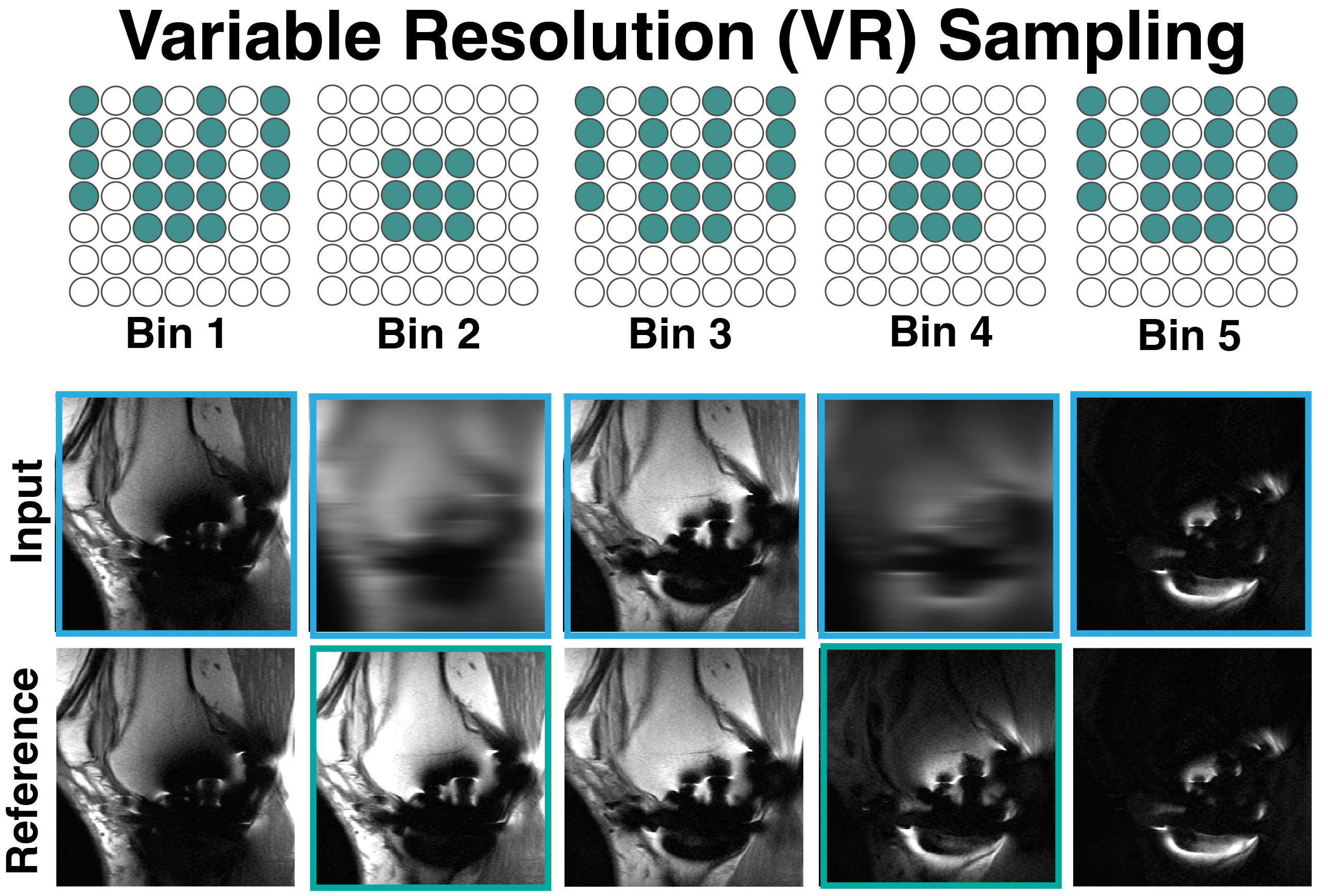}}
\caption{Variable resolution (VR) sampling for faster multi-spectral imaging near metal. In odd numbered bins, conventional parallel imaging with partial Fourier sampling is used. In even numbered bins, however, it is proposed to only acquire the ACS data in the $k_y-k_z$ plane to nearly halve the scan time. This produces very low resolution images that will be used to predict the full-resolution images using a deep learning model. \label{fig:vr}}
\end{figure}

\section{Methods}\label{sec:methods}

\subsection{Datasets} 

Proton density-weighted 1.5T MSI data from the knee of 67 subjects with implanted metal hardware were available for this retrospective study. The data were acquired with 24 spectral bins, a full-width half-maximum RF bandwidth of 2.25 kHz, bin spacing of 1 kHz, an in-plane matrix size of 384x256, and a number of slices ranging between 24 and 32.  These datasets were collected under a research registration protocol approved by the local IRB.  

Images from the first concatenation of bins (i.e., odd bins) were reconstructed using a vendor-provided reconstruction tool (Orchestra, GE HealthCare, Waukesha, WI) which included ARC parallel imaging\cite{brau2007new} and homodyne partial Fourier\cite{noll1991homodyne} processing. Even bins from the second concatenation were retrospectively subsampled to include only the 16x16 center of k-space along the phase encoded dimensions. Low resolution images were reconstructed from the Gaussian window-apodized ACS data with an inverse fast Fourier transform and root-sum-of-squares coil combination. All images were resampled to a 512x512 in-plane matrix size. The number of subjects in the training, validation, and testing datasets were 42, 15, and 10, respectively. 

\subsection{Network Architecture}

The desired outputs of a deep learning model are full-resolution in-plane images in the 12 spectral bins that were reconstructed using only the ACS data. The input to the network are the in-plane images from all 24 available VR bins, a subset of which are depicted in Figure \ref{fig:vr}. A multi-channel 2D U-Net architecture\cite{ronneberger2015u} with a kernel size of 3x3, and five encoding and decoding layers was used to predict the full-resolution bin images from the VR input. The in-plane input to the network was normalized through subtraction of the mean signal and division by the standard deviation. The model was trained using an Adam optimizer for 50 epochs with a learning rate of $2\times 10^{-4}$, a mini-batch size of 4 slices, and a mean squared error loss function. Training time took approximately 8 hours on a GPU with 12 GB of memory (Titan V, NVIDIA, Santa Clara, CA). 

\begin{figure}[ht]
\centerline{\includegraphics[width=15pc]{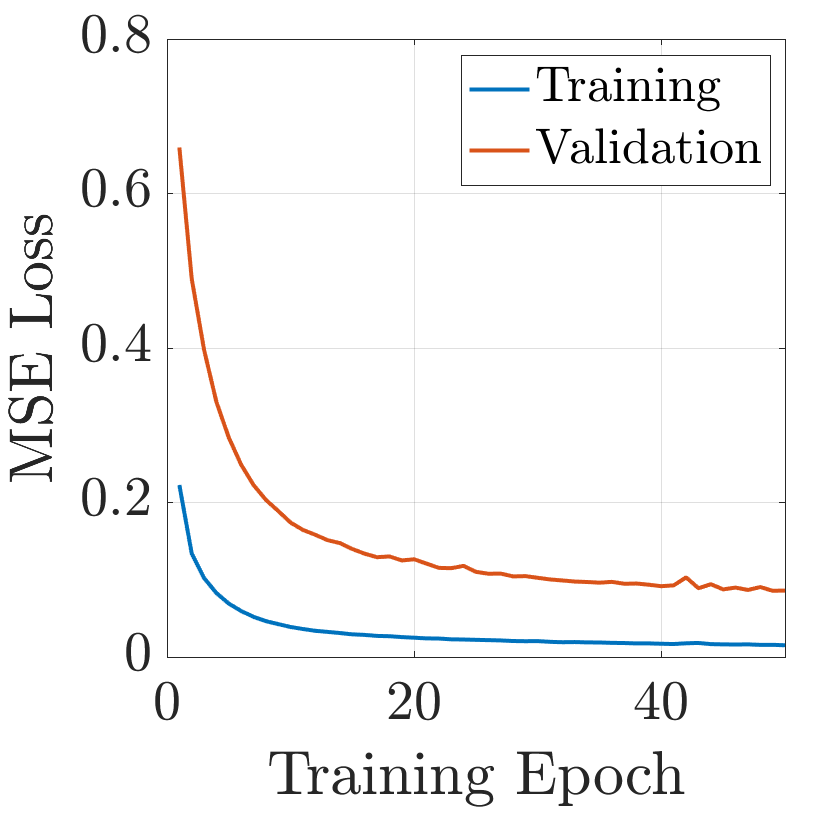}}
\caption{Training and validation dataset loss curves for each epoch during the training of the U-Net for interpolating the VR MSI datasets. \label{fig:loss}}
\end{figure}

\subsection{Analysis} 

The trained network was used to predict full-resolution spectral bin images in all slices for all 10 test subjects. Based on a field map estimated from the magnitude bin images, each individual bin volume was corrected for bin-shift blurring prior to being combined via a root-sum-of-squares operation\cite{koch2011imaging}. Composite bin-combined images were created for the reference data, the unprocessed VR data, and with the data predicted by the deep learning model. Image quality was assessed quantitatively via calculation of structural similarity\cite{wang2004image} (SSIM) and qualitatively via visual assessment. A Gaussian function of the form $a e^{-(\delta - \mathbf{f})^2/(2 \sigma ^2)}$ was fit to the magnitude of the bin images across the spectral domain using a Levenberg-Marquardt non-linear least squares algorithm for the reference, VR, and deep learning-processed datasets. Here, $\mathbf{f}$ is a vector containing the center frequency for each bin. A Wilcoxon rank-sum test was used within a femur region of interest in each subject to test to see whether the reference and U-Net-predicted parameters of the Gaussian fit can be assumed to be from the same distribution.

\section{Results}

\begin{figure*}[h]
\centerline{\includegraphics[width=40pc]{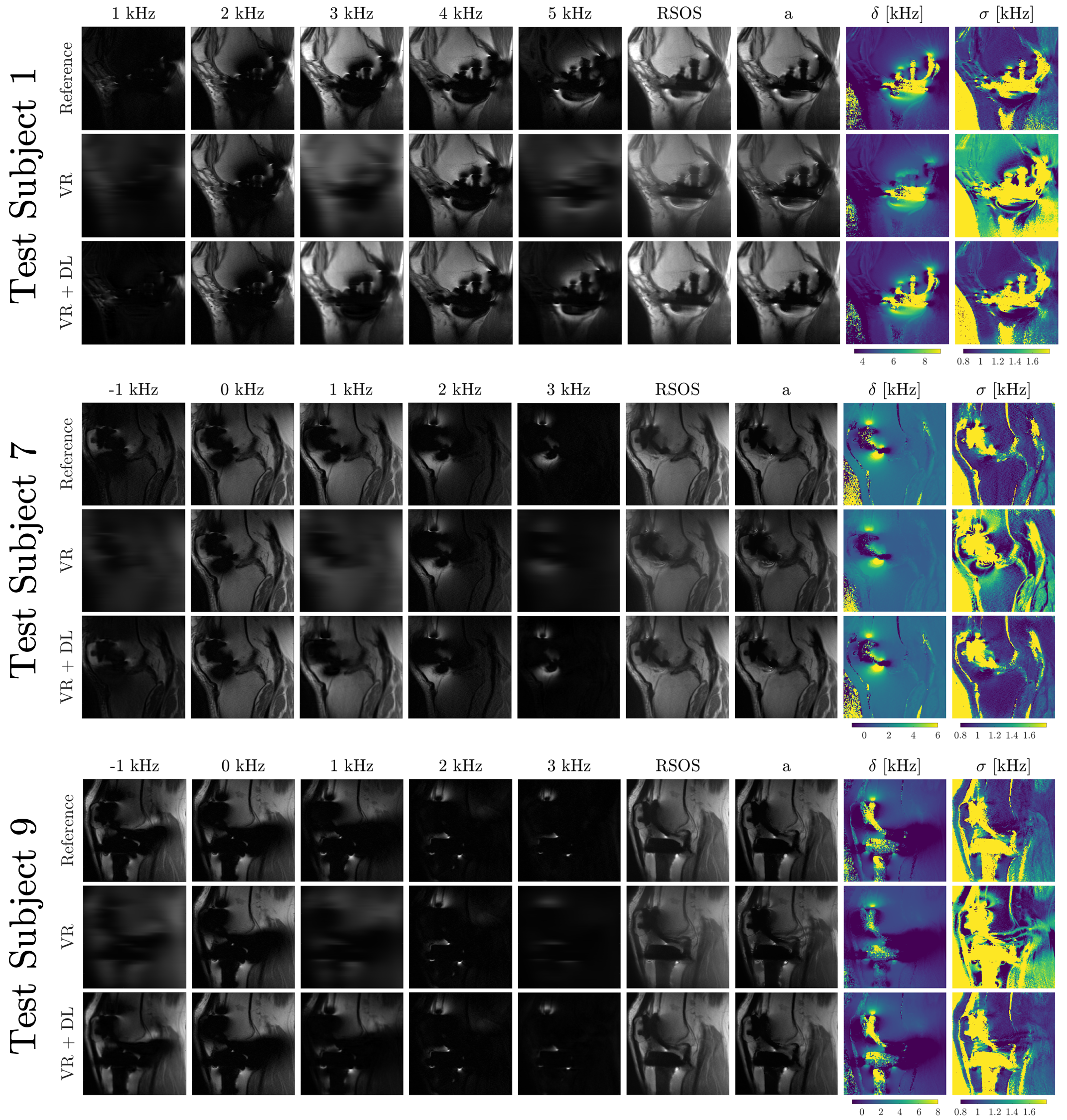}}
\caption{In vivo results in three of the ten test subjects. Five bin images with consecutive center frequencies are shown. The VR images exhibit low resolution in every other bin since they were reconstructed using only calibration data. Bin-combined (RSOS) images are shown in addition to parameters from the fit of a Gaussian function to the signal magnitude in each voxel. A magnified portion of the full field-of-view is shown to better depict differences in the immediate vicinity of the implant.\label{fig:main_result}}
\end{figure*}

Results from 3 of the 10 test subjects can be seen in Figure \ref{fig:main_result}. These images show the reference, VR, and predicted full-resolution images for the five consecutive spectral bins with the most signal in the slice containing the worst metal artifact. Very little detail can be seen in the VR images from bins reconstructed using only the ACS data. The deep learning predictions in these bins can be seen directly beneath the VR images. While not as sharp as the reference images, the predictions offer drastically improved spatial resolution. Following the distortion correction method described above, the root-sum-of-squares (RSOS) images were calculated from each of the datasets. Visually, the VR+DL RSOS images greatly reduced the amount of stripe artifacts introduced by the VR sampling, and are of similar quality to the reference RSOS images. Un-cropped images from all 10 subjects can be seen in Supporting Information Figures S1 - S10. 

\begin{figure}[ht]
\centerline{\includegraphics[width=15pc]{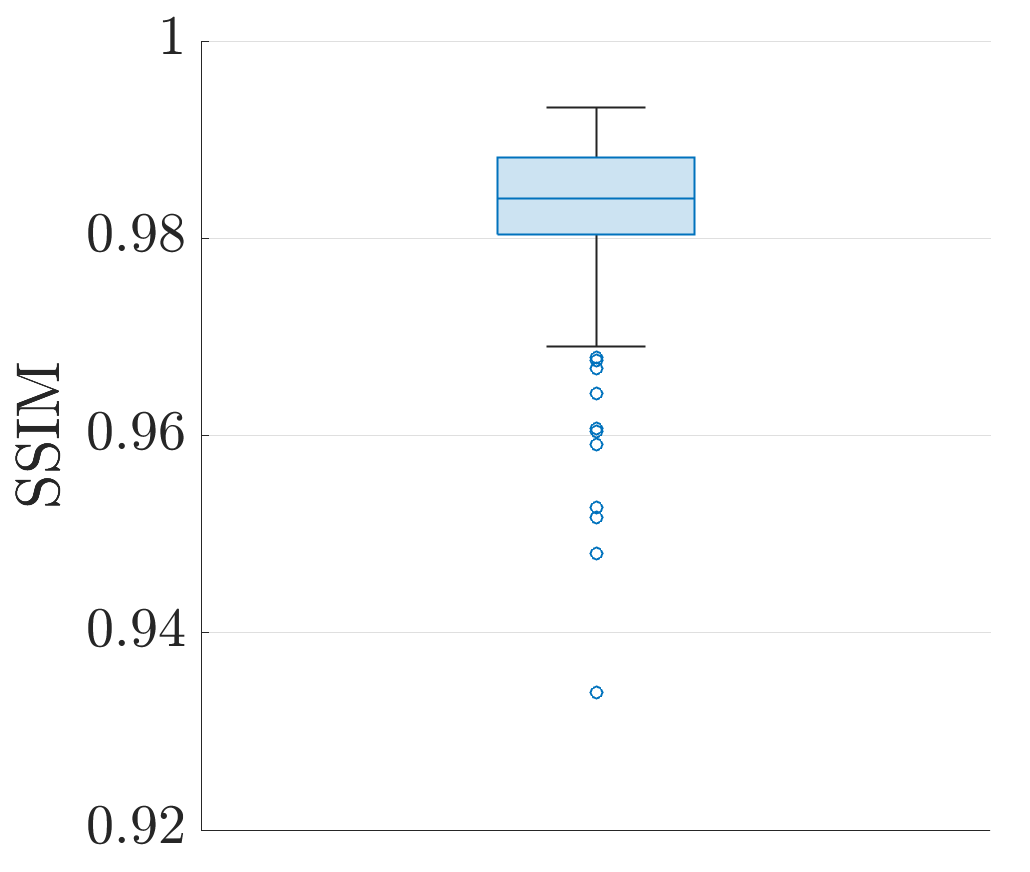}}
\caption{Box and whisker plot showing distribution of structural similarity (SSIM) values between the bin-combined (RSOS) reference and deep learning-predicted images across subjects and slices. \label{fig:ssim}}
\end{figure}

The median and interquartile range (IQR) of the SSIM values across all test subjects and slices was found to be 0.984 and 0.008, respectively, between bin-combined reference and U-Net-predicted images. The distribution of SSIM values can be seen in a box and whisker plot in Figure \ref{fig:ssim}.

\begin{figure}[ht]
\centerline{\includegraphics[width=21pc]{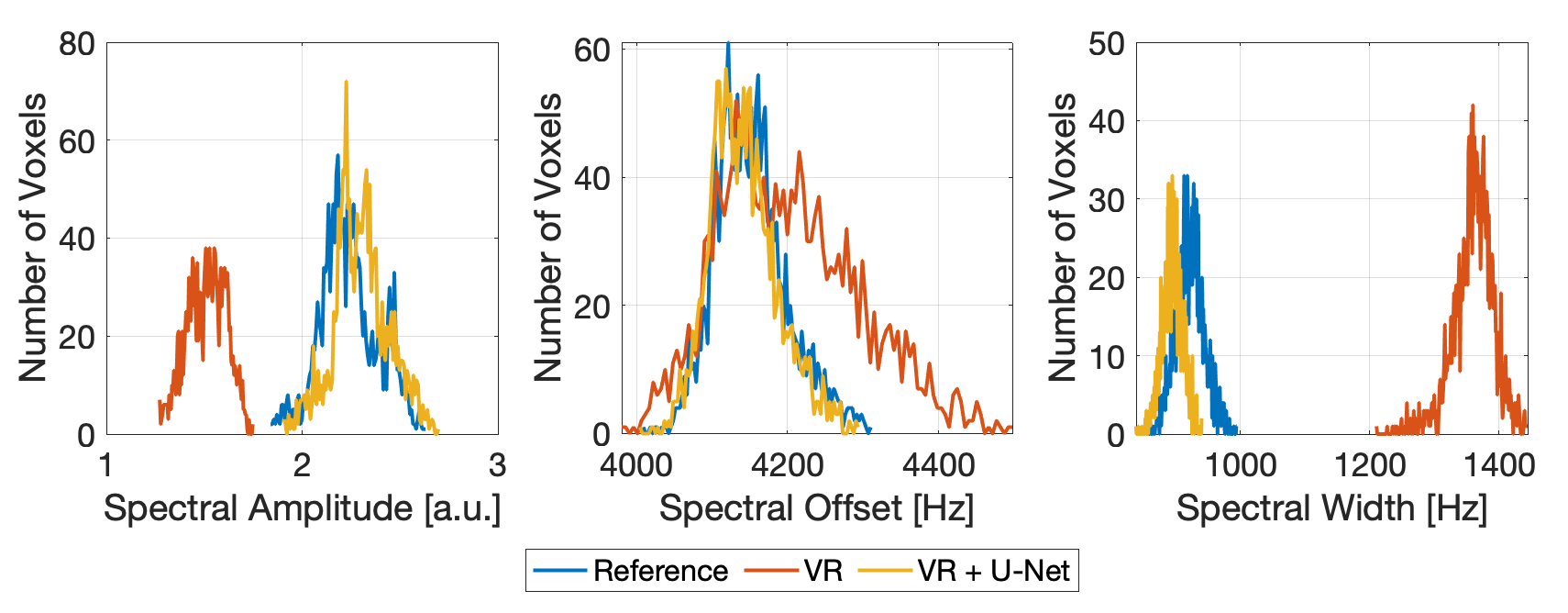}}
\caption{Histograms of spectral amplitude $a$, spectral offset $\delta$, and spectral width $\sigma$ within a femur region of interest for Test Subject 1. The VR sampling introduces bias in the estimated parameters, which the deep learning model largely mitigates. \label{fig:hist}}
\end{figure}

The three right-most columns of Figure \ref{fig:main_result} show the amplitude, center frequency, and standard deviation maps of a Gaussian function fit to the measured spectral profile in each voxel. In all cases, it is readily apparent that the fit of the Gaussian function is highly corrupted by the VR sampling. This is most evident through the severe stripe artifacts in the amplitude ($a$) maps and the bias in the spectral width $\sigma$ maps. The U-Net model helped to produce values of $a$, $\delta$, and $\sigma$ closer to those from the reference (see Figure \ref{fig:hist}), but in all subjects, the fit parameters cannot be assumed to be from the same distribution ($p<0.05$).

\section{Discussion}

In the present work, a variable resolution sampling scheme was proposed in which scan time is nearly halved in multi-spectral imaging near metal. In spectral bin images wherein only the ACS data were acquired, a U-Net deep learning model can be used to predict full-resolution images given VR images across all spectral bins as an input.

While favorable SSIM values were observed (0.984), blurring is apparent in the predicted bin-combined images compared with the reference images. A number of approaches may be used to alleviate this issue. First, inclusion of several slices as inputs to the model may help. The slice resolution is limited by the number of phase encoding lines sampled along this dimension (16 in the present work). Therefore, information about the current slice is contained in neighboring slices as well, which can be exploited by the deep learning model. Further, more advanced architectures incorporating self attention\cite{cao2023swin} or generative-adversarial loss\cite{goodfellow2020generative} may improve image sharpness, though more training data may be required in these cases. 

This study was a retrospective study. k-Space data from conventional MSI were truncated to only the ACS region. This facilitated a direct comparison with reference methods, but introduces minor discrepancies with future prospective studies. Namely, prospectively accelerated MSI datasets will have a different view ordering within each echo train that will change the image resolution due to $T_2$ decay\cite{hennig1986rare}. This is not expected to greatly impact performance since apodization is applied to reduce Gibbs ringing artifacts, but it is a topic that will be investigated in planned prospective studies. 

Overall, this method has the potential to drastically reduce scan time for scanning near metallic implants. A rigorous analysis of the diagnostic image quality obtained with this approach is to be performed in a reader study. 

\bibliography{MRM-AMA}%
\vfill\pagebreak

\clearpage
\section*{Supporting information}
The following supporting information is available as part of the online article:

\vskip\baselineskip\noindent
\textbf{Figure S1.}
{Caption.}

\noindent
\textbf{Figure S2.}
{Caption.}


\end{document}